\documentclass{article}
\usepackage[T1]{fontenc} 
\usepackage[utf8]{inputenc} 
\usepackage[lbd]{ismir} 
\usepackage{amsmath,cite,url}
\usepackage{graphicx,calc} 
\usepackage{color}
\usepackage{xurl}

\usepackage{pgf}
\usepackage{tikz}
\usetikzlibrary[decorations] 
\usetikzlibrary{backgrounds}
\usetikzlibrary{trees}
\usetikzlibrary{arrows,positioning}

\usepackage{xspace}

\def\<#1>{\langle #1 \rangle}

\makeatletter
\newcommand\footnoteref[1]{\protected@xdef\@thefnmark{\ref{#1}}\@footnotemark}
\makeatother

\def\eg{e.g.\xspace}

\def\etc{etc\xspace}

\def\Bar{\mathsf{bar}}

\def\Fork{\mathsf{fork}}
\def\Div{\mathsf{div}}
\def\Deco{\mathsf{deco}}
\def\Orn{\mathsf{orn}}

\def\DO{\scriptstyle\mathsf{C}}
\def\RE{\scriptstyle\mathsf{D}}
\def\MI{\scriptstyle\mathsf{E}}
\def\FA{\scriptstyle\mathsf{F}}
\def\SOL{\scriptstyle\mathsf{G}}
\def\LA{\scriptstyle\mathsf{A}}

\title{TSM Diff: A new parallel-tree-traversal algorithm\\ for the comparison of music score files}

\twoauthors
  {Xiaobo Wang} {INALCO, Paris, France \\ {\tt sibelwang17@gmail.com}}
  {Florent Jacquemard} {Inria Paris, France \\ {\tt florent.jacquemard@inria.fr}}

\def\authorname{X. Wang and F. Jacquemard}

\usepackage[bookmarks=false,pdfauthor={\authorname},hidelinks]{hyperref}

\begin{document}

\maketitle
\begin{abstract}
Music score files are more difficult to compare than their text-based equivalents, 
as their content is more complex and organised into hierarchical structures.
This explains the lack of tools for processing music scores
similar to the widely used old Unix \emph{diff} command~\cite{Miller85diff1}.

We propose a new algorithm and tool for comparing digital music scores in various encodings, 
such as  MusicXML or MEI, based on a 
graph-structured abstract representation called Tree Score Model (TSM).
\end{abstract}

\section{Intermediate Representation}\label{sec:TSM}\label{sec:IR}
Our approach builds on two previous tools: 
ScoreDiff~\cite{Foscarin19dlfm} and its fork MusicDiff~\cite{Martinez25benchmark}. 
The most crucial difference is a change in the intermediate representation of score content 
used for comparison:
the two above tools are based on unstructured Music21 streams~\cite{Cuthbert10music21}, 
whereas we use a new representation TSM structured as labelled directed acyclic graphs (DAGs), 
inspired by rhythm trees~\cite{Agon02RRR}.
These DAGs encode the metrical and rhythmic structure of a score, in the sense that, 
roughly speaking, every branching corresponds to the division of a musical time interval.
More specifically, the nodes of the DAG are labelled with basic events (notes, rests, etc.) for leaf nodes, 
and, for inner nodes, with functions that associate several sub-intervals (one per child) 
with a time interval. In this way, we can associate a time interval with each node of the DAG. 

\begin{figure}
\centerline{\includegraphics[width=0.8\columnwidth]{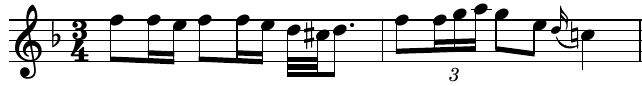}}
\centerline{
\scalebox{0.85}{
\(
\begin{tikzpicture} [baseline,-,thick]
\tikzstyle{level 1}=[level distance=10mm, sibling distance=54mm]
\tikzstyle{level 2}=[level distance=9mm, sibling distance=21mm]
%
\node {$\Bar$} 
child { node {$\Fork$}  
child[level 3/.style={sibling distance=17mm},
        level 4/.style={sibling distance=7mm},
        level 5/.style={sibling distance=7mm},
        level 6/.style={sibling distance=6mm}] { node {$\Div_3$}  
  child { [sibling distance=16mm] node {$\Div_2$} 
    child { node {$\FA_5$} }
    child { node {$\Div_2$} 
     child { node {$\FA_5$} }
     child { node {$\MI_5$} } } }
  child { node {$\Div_2$} 
    child { node {$\FA_5$} }
    child { node {$\Div_2$} 
      child { node {$\FA_5$} }
      child { node {$\MI_5$} } } }
  child { node(share2) [xshift=9mm] {$\Div_2$} 
    child { node {$\Div_2$} 
      child { node {$\Div_2$} 
         child { node {$\RE_5$} }
         child { node {$\DO_5\sharp$} } }
      child { node(share3) {$\RE_5$} } }
    child[missing] { node {$\RE_5$} } } } }
%
child { node {$\Bar$}
child { node [xshift=8mm] {$\Fork$}  
  child[level 4/.style={sibling distance=14mm},
          level 5/.style={sibling distance=7mm},
          level 6/.style={sibling distance=5mm},
          level 7/.style={sibling distance=5mm}]  { node {$\Div_3$}
    child { node {$\Div_2$} 
       child { node {$\FA_5$} }
       child { node {$\Div_3$}
         child { node {$\FA_5$} } 
         child { node {$\SOL_5$} } 
         child { node {$\LA_5$} } } }
    child { node {$\Div_2$} 
       child { node {$\SOL_5$} }
       child { node {$\MI_5$} } }
    child { node {$\Deco$} 
      child { node {$\Orn$} 
        child { node {$\RE_5$} } }
      child { node {$\DO_5$} } } } }
%
%
child { node {$\mathsf{end}$} } };  
\draw (share2) to[bend left=40] (share3);	
\end{tikzpicture}
\)
} 
} 
\caption{TSM encoding of two monophonic 3/4 bars.}
\label{fig:TSM-example}
\end{figure}

For instance, Figure~\ref{fig:TSM-example} contains the DAG encoding of two bars of 
a monophonic reduction of Polonaise 
BWV Anh~II~128.
The root function $\Bar$ divides the top time interval $[0, 2[$,
covering the whole extract (durations are expressed in number of bars),
into the interval $[0, 1[$ of duration 1 bar and
$[1, 2[$ for the rest of the excerpt. 
The function $\Fork$ duplicates
into several concurrent voices of same duration -- 
in this example, for the sake of simplicity, we consider only~1 voice. 
The functions $\Div_2$ and $\Div_3$ partition one time interval
into 2, respectively 3, sub-intervals of equal duration.
They correspond to the regular division of the bar's content into beats, sub-beats \etc, 
as defined by the time signature, or to \textit{irregular} tuplets like the triplet of 16th notes in the second bar
of Figure~\ref{fig:TSM-example}.
The function $\Deco$ at the end of the second bar splits 
the interval $[\frac{5}{3}, 2[$, spanning the last beat, into 
the empty $[\frac{5}{3}, \frac{5}{3}[$,
for the ornament $\Orn(\RE_5)$ made of a single grace-note, 
and $[\frac{5}{3}, 2[$,
for the decorated note~$\DO_5$.
%
Finally, 
the last node in the first bar, labelled with $\RE_5$,
is shared by its parent and grandparent, 
both of which are labelled with~$\Div_2$. 
This means that the interval associated with $\RE_5$ 
is the sum of the two intervals 
for the second child of each of these two ancestor nodes, 
with respective durations of~$\frac{1}{12}$ (sixteenth note) and~$\frac{1}{6}$ (eighth note), 
namely $[\frac{3}{4}, \frac{5}{6}[ \cup [\frac{5}{6}, 1[ = [\frac{3}{4}, 1[$.
This sum corresponds to extending an eighth note with one dot.

It is important to note that, since durations are encoded by the structure of the DAG itself, leaf nodes do not need duration attributes such as note values.
The only attribute of the note leaves in the example of Figure~\ref{fig:TSM-example} is their pitch.
Notational information, such as dynamics, articulations, fermatas, ornaments \etc, can also be attached to the nodes, 
without changing the associated time intervals.

\section{Comparison Algorithm}\label{sec:algo}\label{sec:diff}
Given two scores, after conversion into TSM DAGs, our procedure calculates jointly: 
\begin{itemize}
\item A list of differences (\emph{diffs}); 
\item A time distance: cumulated durations of time intervals where scores differ;
\item An edit distance, called surface distance, based on notational attributes not related to time. 
\end{itemize}
The two latter measures are normalised and combined into a similarity distance between scores.
Each {diff} consists of the nodes involved in the DAGs in comparison,
with, for each score, 0, 1 or several nodes.
The last case occurs when meeting shared nodes, for resynchronizing,
so that every \emph{diff} covers the same time interval in both scores. 
%
The relevant information reported to the user
is the labels of these nodes and the corresponding time intervals. 
%
Moreover, a \emph{diff code} 
can optionally be provided 
to help interpreting the meaning of the {diff}.
In the examples and demos below, we use 
\emph{insert}, \emph{delete}, \emph{fragment} and \emph{consolidate},
\emph{replace}, and some specialised 
variants\footnote{see \url{https://tsm-diff-lbd-7a7e2a.gitlab.io} for more details on the diff codes in the demos.}.

The computation proceeds first bar-wise, aligning equal bars with a
\emph{Longest Common Subsequence} (LCS) algorithm~\cite{hunt1977fast}, 
and then \emph{matching} the voices inside differing bars, with a dedicated algorithm.
The next crucial step is the comparison of the content of two voices, 
performed by a \emph{recursive parallel descent} into the two sub-DAG representations.
The descent stops as soon as the labels diverge.
In this case, a new difference is added to the list, the duration of the involved intervals is added to the time distance, 
and the surface distance is also updated with 
an edit distance between the sequence of attributes not related to time
in the leaves under these nodes.

\begin{figure}
\begin{center}
\raisebox{1ex-\height}{\includegraphics[width=0.40\columnwidth]{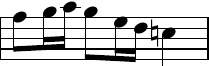}}
\quad
\raisebox{1ex-\height}{%
\scalebox{0.89}{
\(
\begin{tikzpicture} [baseline,-,thick]
\tikzstyle{level 1}=[level distance=10mm, sibling distance=15mm]
\tikzstyle{level 2}=[level distance=9mm, sibling distance=7mm]
\tikzstyle{level 3}=[level distance=9mm,sibling distance=6mm]
\tikzstyle{level 4}=[level distance=9mm,sibling distance=6mm]
\node {$\Div_3$}  
    child { node {$\Div_2$} 
       child { node {$\FA_5$} }
       child { node {$\Div_2$}
         child { node {$\SOL_5$} } 
         child { node {$\LA_5$} } } }
    child { node {$\Div_2$} 
       child { node {$\SOL_5$} }
       child { node {$\Div_2$}
         child { node {$\MI_5$} } 
         child { node {$\RE_5$} } } }
  child { node [xshift=-2mm] {$\DO_5$} };
\end{tikzpicture}\)
} 
} 
\end{center}
\caption{Variant for the content of Figure~\ref{fig:TSM-example} second bar.}
\label{fig:variant}
\end{figure}

For instance, a parallel descent into the content of the second bar in Figure~\ref{fig:TSM-example}
(sub-DAG below the second $\Fork$)
and the variant depicted in Figure~\ref{fig:variant}, will stop 3 times, once in each beat:
in beat~1, the diverging nodes are 
$\Div_3$ in Figure~\ref{fig:TSM-example} and $\Div_2$ in Figure~\ref{fig:variant},
this produces a rhythm-change diff, corresponding to a \emph{replace} operation between the two division nodes. 
The time distance is twice~$\frac{1}{6}$ (8th note) and the surface distance is~1 (deletion of $\FA_5$);
in beat~2, 
we identify a \emph{fragmentation}, in the sense of~\cite{Mongeau90comparison}:
the replacement of a single note $\MI_5$ by several notes $\Div_2(\MI_5, \RE_5)$, 
while preserving the total duration,
and in beat~3, the pair made of $\Deco$ and $\Div_2$
corresponds to the \emph{deletion} of an ornament (grace note $\RE_5$).
In this last case, we continue the recursive descent in the right child $\DO_5$ of $\Deco$ but no other difference is found.
%
When two matched nodes have annotation lists, 
we compare them with an edit distance, added to the surface distance, 
and yielding new insert, delete or replace operations in the diff list.

\section{Applications, Demos}\label{sec:demos}\label{sec:applications}
\noindent{\bf Implementation.}
The above 
procedure has been implemented in {C++}, 
on top of a {C++} library dedicated to TSM. 
Moreover, Python binding and import functions are written to
convert MusicXML or MEI scores into TSM DAGs, 
using the Music~21 toolkit~\cite{Cuthbert10music21} for XML parsing.

\noindent{\bf Results.}
As a proof of concept
we propose two demos: 
$(i)$~5 movements from piano sonatas taken from~\cite{Hentschel21MozartDCML}, 
each of which being compared with 3 versions produced manually (with increasing numbers of edits);  
$(ii)$~10~comparisons between monophonic soli transcriptions of 5 different performers of the jazz standard \emph{All the Things You Are},
taken from~\cite{Foster21FiloSax}.
%
%
Table~\ref{tab:demo-results} 
presents some results\footnote{\label{note:results}see \url{https://tsm-diff-lbd-7a7e2a.gitlab.io/demo.html} for complete results.}
for each demo.
%
In all cases, the TSM diff core comparison time 
is 1.7ms to 28.3ms (average~5ms), 
and the overall time (incl. Music21 parsing and TSM construction) is at most~$\sim$4.2s.

\noindent{\bf Applications.}
The main purpose of our approach to comparing musical scores is 
\emph{version management} in the context of editing music notation, 
or the \emph{creation and maintenance of corpora} for musicological studies, 
for example to identify duplicate or similar scores.
It could also be applied \eg to \emph{comparative corpus studies} in musicology, 
see \eg~\cite{Nachtwey25}.

The similarity distance 
could also be useful for the \emph{evaluation} of 
transcription~\cite{Cogliati17,Suzuki21ScoreTransformer,Nakamura18icassp,Hiramatsu2021joint,McLeodSteedman18MV2H}
or OMR~\cite{Martinez25benchmark} 
procedures, 
for comparing the outcome 
with a reference XML score. 

\begin{table}
\centering
\begin{tabular}{l|rrr|rr}
\hline
 & \multicolumn{3}{|c}{TSM-diff} & \multicolumn{2}{|c}{MusicDiff}\\
\hline
{\small demo} & {\small diffs} & {\small simil.} & {\small time} & {\small diffs} & {\small time}\\
\hline
$\mathrm{\scriptstyle DCML}_1 \mathrm{\scriptstyle min}$ & 1 & 0.999 & 2.59s & 1 & 2.23s \\
\multicolumn{1}{r|}{$\mathrm{\scriptstyle max\ diffs}$} & 19 & 0.992 & 4.20s & 54 & 3.72s \\
\hline
$\mathrm{\scriptstyle DCML}_2 \mathrm{\scriptstyle min}$ & 73 & 0.958 & 1.65s & 320 & 5.41s \\
\multicolumn{1}{r|}{$\mathrm{\scriptstyle max\ diffs}$} & 304 & 0.796 & 2.67s & 772 & 43.94s\\
\hline
$\mathrm{\scriptstyle DCML}_3 \mathrm{\scriptstyle min}$ & 432 & 0.763 & 4.08s & 1166 & 24.15s \\
\multicolumn{1}{r|}{$\mathrm{\scriptstyle max\ diffs}$} & 633 & 0.672 & 2.68s & 1322 & 21.72s \\
\hline
$\mathrm{\scriptstyle FiloSax}\,\mathrm{\scriptstyle min}$ & 323 & 0.359 & 1.26s & 858 & 333.40s\\
\multicolumn{1}{r|}{$\mathrm{\scriptstyle max\ diffs}$} & 645 & 0.378 & 1.50s & -- & $>$ 15m \\
\hline
\end{tabular}
\caption{Partial results for the demos$^{\ref{note:results}}$.}
\label{tab:demo-results}
\end{table}

\section{Final Remarks}
This recursive parallel descent algorithm yields linear time complexity,
ensuring 
high efficiency and robustness of the comparison procedure, 
more so than if it were based solely on quadratic computations 
of edit distances like~\cite{Foscarin19dlfm, Martinez25benchmark}.

Furthermore, our approach clearly separates
the comparison of time-related aspects (which is performed first)
from other aspects,
in accordance with the principle of \emph{disjoint penalties}
for transcription evaluation~\cite{McLeodSteedman18MV2H}.

Finally, the \textit{diff codes} listed above were chosen arbitrarily. 
As they are 
added during post-processing, 
they are in fact independent of the algorithms used for comparison 
and could easily be modified, as required, for other case studies
in MIR or digital musicology.



\section{AI Usage Statement}
No large language models (LLMs) were used either to conduct the research described in this article, 
or to write the article itself, 
with the exception of programming aids, 
as well as editing tools, such as 
spell-checking or grammar suggestions.

\bibliography{references}

%
%
%
%
%

\end{document}